\documentclass[preprint,preprintnumbers,showpacs]{revtex4}
\usepackage{amsmath}
\usepackage{amssymb}
\usepackage{graphicx}
\usepackage{dcolumn}
\usepackage{bm}

\begin{document}

\title{Quantum decoherence of spin states in an electric-field controllable single molecular magnet}

\author{Xiang Hao}

\author{Chen Liu}

\affiliation{Department of Physics, School of Mathematics and
Physics, Suzhou University of Science and Technology, Suzhou,
Jiangsu 215009, People's Republic of China}

\author{Shiqun Zhu}
\altaffiliation{Corresponding author: szhu@suda.edu.cn}

\affiliation{School of Physical Science and Technology, Soochow
University, Suzhou, Jiangsu 215006, People's Republic of China}

\begin{abstract}

The time evolution of low energy spin states of a single molecular magnet in a local electric field is investigated. The decoherence of the driven single molecular magnet weakly coupled to a thermal bosonic environment is analyzed by the second-order time-convolutionless non-Markovian master equation. If the characteristic time of the system is much smaller than the correlation time of the environment, the analytical expression of the reduced density matrix of the system is obtained. The non-Markovian dynamics of the spin states at low temperatures is induced by the memory effects in the decay rates. The non-Markovian oscillation of the Bloch vector is presented. The quantum decoherence can be effectively restrained with the help of the reasonable manipulation of the environment spectral density function and local electric field. The influence of the dissipation on the pointer states are investigated by the von Neumann entropy. The pointer states can be selected by the environment.

\end{abstract}

\maketitle

\section{INTRODUCTION}

The unavoidable interactions of all open quantum systems with environments often result in the dissipation and decoherence \cite{Breuer01, Weiss99}. Due to the exchange of energy and information between the system and the environment, the non-Markovian dynamics of quantum states always occur in the realistic experimental systems \cite{Hoeppe, Guo, Kuhr}. Recently, much attentions have been paid to the control of the decoherence of many-body quantum systems \cite{Carle, Ardavan} such as spin clusters and single molecular magnets \cite{Gatteschi}. As a class of systems with rich quantum properties, single molecular magnets at low energies can serve as a large-spin system or a collection of interacting spins \cite{Friedman, Thomas, Wernsdorfer}. These solid quantum spin systems are considered as promising carriers of quantum information \cite{Leuenberger, Lehmann}. Single molecular magnets with antiferromagnetic spin couplings can provide low-energy states for performing quantum logic gates \cite{Geogeot, Timco, Candini}. Quantum decoherence is manifested when single molecular magnets are coupled to a spin bath or thermal reservoir \cite{Carle, Ardavan, Szallas, Morello, Prokofev, Coish}. The dissipation and decoherence always depend on the properties of the environment which can be described by a certain spectral density function \cite{Biercuk, Maniscalco, Haikka, Goan}. Therefore, a reasonable quantum manipulation method is necessary. At present, chemical manipulation can offer an efficient way to engineer intermolecular couplings and allow for interactions between qubits \cite{Troiani}. The decoherence from the chemical control cannot be easily eliminated because of the permanent interactions with the surrounding \cite{Szallas}. Simultaneously, the most straightforward and conventional way is to adopt an external magnetic field produced by electron spin resonance pulses \cite{Ardavan}. Although the decoherence of single molecular magnets can be suppressed by strong magnetic fields, it is preferable to apply electric fields that are controllable and suitable on very small spatial and temporal scales. It is possible to apply time-dependent strong electric field close to single molecular magnet via a scanning tunnel microscopic tip \cite{Trif, Hirji06}. It has been found out that an electric field can be coupled to low-energy spin states of different chirality due to the absence of spin inversion symmetry in some single molecular magnets, such as $Cu_{3}$ \cite{Trif} , $V_{15}$ \cite{Chiorescu}, $Co_{3}$ \cite{Juan}, $Dy_{3}$ \cite{Luzon} and $Mn_{12}$ \cite{Friedman, Thomas}, etc. The effective spin electric coupling relies on the detailed structure of single molecular magnets at low energies \cite{Trif10}. Moreover, the use of microwave cavities can contribute to the indirect generation of fully controllable and long-range interaction between any two molecular magnets. This scheme based on electric-field local control can open up the possibility of scalable solid quantum information processing. Thus, the control of the decoherence of single molecular magnets driven by local electric field in low-temperature environments needs to be further studied.

This paper is organized as follows. In Sec. II, the time evolution of low-energy spin states in an electrically driven single molecular magnet without the spin inversion symmetry is studied. Using the general low-energy spin Hamiltonian with effective spin-electric coupling, the time-convolutionless non-Markovian master equation is derived under the assumption of weak interactions with a thermal bosonic environment. If the characteristic time scale of the system is much shorter than the correlation time of the environment, the reduced density matrix of the low-energy spin states can be obtained analytically. As an example, the effects of the low temperatures on the decaying rates of $Cu_{3}$ are calculated. In Sec. III, the non-Markovian decoherence of spin states is described by means of the Bloch vector. The traditional Lorentzian spectral density function of the environment is considered. The selection of the pointer state measured by the von Neumann entropy is also investigated. Finally, a simple discussion concludes the paper.

\section{MODEL OF DECOHERENCE IN THERMAL BOSONIC ENVIRONMENT}

Molecular magnets have clear features of coherent behavior and a variety of effective low-energy spin Hamiltonian is used for encoding qubits and implementing spin-based quantum computation. A local electric field $\vec{\epsilon}(t)$ can be coupled to low-energy spin states of opposite chirality because of the effective spin electric interactions \cite{Trif10}. It is shown that both spin-orbit interactions and the absence of spin inversion symmetry can induce the the electric dipole matrix element $\vec{d}$ which is an important quantity in the effective spin-electric coupling. The strength of spin-electric coupling can be calculated by means of {\it ab initio} methods. For a single molecular magnet with an effective spin-electric coupling, the effective low-energy spin Hamiltonian is written as
\begin{equation}
\label{eq:1}
H_{eff}=H_{0}+H_{\epsilon},
\end{equation}
where $H_{0}$ is the low-energy spin Hamiltonian without the electric field and the effective spin-electric coupling is given by $H_{\epsilon}=\vec{d}\cdot \vec{\epsilon}=\sum_{i}e\vec{r}_{i}\cdot \vec{\epsilon}$. Here $e$ is the electron charge and $\vec{r}_{i}$ denotes the coordinate of the $i$th electron in the spin structure of single molecular magnet. In the following discussion, a simple case of triangular spin-$\frac 12$ molecular magnet $Cu_{3}$ \cite{Trif} is considered. In the low-energy Hilbert subspace spanned by two opposite chiral spin states $\{ |C=\pm 1, M=\frac 12\rangle\}$, the effective spin Hamiltonian in the presence of effective spin-orbit interaction $\omega_{so}$ and planar electric field $\vec{\epsilon}(t)$ can be expressed as
\begin{equation}
\label{eq:2}
H_{eff}=\frac 12 \omega_{so} C_{z}+d\cdot \epsilon[e^{-i(\omega t+\alpha)}C_{+}+e^{i(\omega t+\alpha)}C_{-}].
\end{equation}
The chirality operator is $C_{z}=\frac {4}{\sqrt{3}}\vec{s}_{1}\cdot (\vec{s}_{2}\times \vec{s}_{3})$ where $\vec{s}_{i}$ denotes the spin operator of $i$th site. The operators $C_{\pm}$ reversing the chirality of the spin states satisfy that $C_{\pm}|C=\mp 1,M\rangle=|C=\pm 1,M\rangle$ and $C_{\pm}|C=\pm 1,M\rangle=0$. The chirality operator behaves similar to the spin operator in the chiral space. Here $|C=\pm 1,M\rangle$ are the simultaneous eigenvectors of $C_{z}$ and total spin operator $S_{z}=\sum_{i}s_{i}^{z}$ with the corresponding eigenvalue $C$ and $M$. The ground state $M=\frac 12$ is taken into account. The parameter $\omega$ represents the frequency of the electric field. $d=|\vec{d}|$ describes the strength of the electric dipole. With the variation of the initial angle between the field $\vec{\epsilon}$ and the vector $\vec{r}_{12}$ from site $1$ to site $2$, we can reasonably adjust the angle $\alpha=0$ \cite{Trif, Trif10}.

A thermal bath can be described by an infinite chain of quantum harmonic oscillators. In the rotating frame with the electric field frequency $\omega$, the total Hamiltonian of the open system coupled to the environment can be written as
\begin{equation}
\label{eq:3}
H=H_{eff}+H_{E}+H_{I}.
\end{equation}
The effective spin Hamiltonian of the molecular magnet is transformed to
\begin{equation}
\label{eq:4}
H_{eff}=\frac 12 (\Delta_{so}C_{z}+d\cdot \epsilon C_{x}),
\end{equation}
where $\Delta_{so}=\omega_{so}-\omega$ and the chirality operator $C_{x}=\frac 12(C_{+}+C_{-})$. The Hamiltonian of the thermal environment is given by
\begin{equation}
\label{eq:5}
H_{E}=\sum_{j}\omega_{j}b^{\dag}_{j}b_{j},
\end{equation}
where $b_{j}$ and $b^{\dag}_{j}$ are the annihilation and creation operator in the Hilbert space of bosonic environment. The last term in Eq. (3) denotes the weak interaction between the system and the environment, and can be written as
\begin{equation}
\label{eq:6}
H_{I}=\sum_{j}(g_{j}e^{-i\omega t}b^{\dag}_{j} C_{-}+g^{\ast}_{j}e^{i\omega t}b_{j} C_{+}).
\end{equation}
Here the rotating wave approximation is adopted. The weak coupling $|g_{j}|=d\cdot \epsilon_{j}$ where $\epsilon_{j}$ represents the magnitude of the electromagnetic field of the $j$th mode with the frequency $\omega_{j}$. For the convenience, the Planck constant $\hbar$ is set to be one. In the interaction representation, the decoherence of the low-energy spin state $\rho(t)$ can be approximately given by the second-order time-convolutionless master equation,
\begin{equation}
\label{eq:7}
\frac {d\rho(t)}{dt}=-\int_{0}^{t}dt_{1} \mathrm{Tr}_{E}[H^{\prime}_{I}(t),[H^{\prime}_{I}(t_1),\rho(t)\otimes\rho_{E}]],
\end{equation}
where $H^{\prime}_{I}(t)=e^{it(H_{eff}+H_{E})}H_{I}e^{-it(H_{eff}+H_{E})}$. The notation $\mathrm{Tr}_{E}$ is the partial trace over the freedom of the environment. It is assumed that the initial product state of the total system is $\rho_{tot}(0)=\rho(0)\otimes \rho_{E}$ where $\rho_{E}=\exp(-H_{E}/\kappa_{B}T)/\mathrm{Tr}[\exp(-H_{E}/\kappa_{B}T)]$ is the thermal equilibrium state of the environment and satisfies that $\mathrm{Tr}[H^{\prime}_{I}(t)\rho_{E}]=0$ \cite{Goan}. The low temperature condition of $\kappa_{B}T<\omega_{so}$ is considered here. For the convenience, the Boltzmann constant $\kappa_{B}$ is also assumed to be one.

To simplify the analytical calculation, the effective spin Hamiltonian can be diagonalized as $\bar{H}_{eff}=\frac {\omega_{s}}{2}\bar{C}_{z}$ where $\omega_{s}=\sqrt{\Delta^2_{so}+d^2\epsilon^2}$. The transformed chirality operator is $\bar{C}_{z}=U^{\dag}C_{z}U=|\Uparrow\rangle \langle \Uparrow|-|\Downarrow\rangle \langle \Downarrow|$ where the transformation operation is $U=|\psi_{+}\rangle \langle \Uparrow|+|\psi_{-}\rangle \langle \Downarrow|$. The eigenvector of $H_{eff}$ are $|\psi_{\pm}\rangle=\pm\sqrt{\delta_{\pm}}|C=+1,M=\frac 12\rangle+\sqrt{\delta_{\mp}}|C=-1,M=\frac 12\rangle$. The coefficients $\delta_{\pm}=(\omega_{s}\pm \Delta_{so})/2\omega_{s}$.  Then, the interaction Hamiltonian in this dressed state basis $|\Uparrow(\Downarrow)\rangle$ can be given by
\begin{equation}
\label{eq:8}
\bar{H}^{\prime}_{I}(t)=A^{\dag}(t)\otimes B(t)+A(t)\otimes B^{\dag}(t).
\end{equation}
Here $A(t)=\sum_{j}g_j e^{-i\omega_j t}b_j$ and $B^{\dag}(t)=\delta_0 \bar{C}_z+\delta_{+} e^{i\omega_s t}\bar{C}_{+}-\delta_{-} e^{-i\omega_s t}\bar{C}_{-}$ where $\delta_0=\sqrt{\delta_{+} \delta_{-}}$. The expression of the time-convolutionless master equation in the dressed state basis is obtained as
\begin{equation}
\label{eq:9}
\frac {d\bar{\rho}(t)}{dt}=-i[\bar{H}_{eff}+\bar{H}_{L},\bar{\rho}(t)]+L[\bar{\rho}(t)]+NL[\bar{\rho}(t)].
\end{equation} Where $\bar{H}_{L}=\mathrm{Im}(\Gamma_0-\Gamma^{\prime}_0)\delta^2_{0}\bar{C}^2_{z}+\sum_{q=\pm}\mathrm{Im}(\Gamma_{q}-\Gamma^{\prime}_{q})\delta^2_{q}\bar{C}^{\dag}_{q}\bar{C}_{q}$. The parameters $\Gamma_{q}$ and $\Gamma^{\prime}_{q}(q=0,\pm)$ are determined by
\begin{eqnarray}
\label{eq:10}
\Gamma_{q}&=&\int_{0}^{t}dt_{1}\sum_{j}|g_j|^2\cdot \bar{n}_{j}e^{(\omega_j-\omega-q\omega_s)(t-t_1)} \nonumber \\
\Gamma^{\prime}_{q}&=&\int_{0}^{t}dt_{1}\sum_{j}|g_j|^2\cdot(\bar{n}_{j}+1) e^{(\omega_j-\omega-q\omega_s)(t-t_1)},
\end{eqnarray}
where $\bar{n}_j=(e^{\omega_j/T}-1)^{-1}$ is the mean number for the $j$th mode of the thermal environment at $T$ temperature.
The Lindblad superoperator in Eq. (9) is given by
\begin{equation}
\label{eq:11}
L[\bar{\rho}(t)]=\sum_{m=z,\pm}\gamma_{m}(t)[\bar{C}_{m}\bar{\rho}\bar{C}^{\dag}_{m}-\frac 12\{\bar{C}^{\dag}_{m}\bar{C}_{m}, \bar{\rho} \}],
\end{equation}
where the decay rates are obtained as $\gamma_{z}(t)=2\delta^2_0\mathrm{Re}(\Gamma_0+\Gamma^{\prime}_0)$, $\gamma_{+}(t)=2\delta^{+}_0\mathrm{Re}(\Gamma_{+})+2\delta^{-}_0\mathrm{Re}(\Gamma^{\prime}_{-})$ and $\gamma_{-}(t)=2\delta^{-}_0\mathrm{Re}(\Gamma_{-})+2\delta^{+}_0\mathrm{Re}(\Gamma^{\prime}_{+})$. The notation $\mathrm{Im(Re)}$ denotes imaginary (real) part of a complex parameter.
The last term in Eq. (9) is very complicate,
\begin{widetext}
\begin{eqnarray}
\label{eq:12}
NL[\bar{\rho}(t)]&=&\Gamma_{0}\cdot[\delta_0\delta_{+}(\bar{C}_z\bar{\rho}\bar{C}_{-}-\bar{\rho}\bar{C}_{-}\bar{C}_z)-\delta_0\delta_{-}(\bar{C}_z\bar{\rho}\bar{C}_{+}-\bar{\rho}\bar{C}_{+}\bar{C}_z)]\nonumber \\
&+&\Gamma_{+}\cdot[\delta_0\delta_{+}(\bar{C}_{+}\bar{\rho}\bar{C}_{z}-\bar{\rho}\bar{C}_{z}\bar{C}_{+})-\delta_{+}\delta_{-}(\bar{C}_{+}\bar{\rho}\bar{C}_{+}-\bar{\rho}\bar{C}_{+}\bar{C}_{+})]\nonumber\\
&-&\Gamma_{-}\cdot[\delta_0\delta_{-}(\bar{C}_{-}\bar{\rho}\bar{C}_{z}-\bar{\rho}\bar{C}_{z}\bar{C}_{-})+\delta_{+}\delta_{-}(\bar{C}_{-}\bar{\rho}\bar{C}_{-}-\bar{\rho}\bar{C}_{-}\bar{C}_{-})]\nonumber\\
&+&\Gamma^{\prime}_{0}\cdot[\delta_0\delta_{+}(\bar{C}_{-}\bar{\rho}\bar{C}_{z}-\bar{\rho}\bar{C}_{z}\bar{C}_{-})-\delta_0\delta_{-}(\bar{C}_{+}\bar{\rho}\bar{C}_{z}-\bar{\rho}\bar{C}_{z}\bar{C}_{+})]\nonumber \\
&+&\Gamma^{\prime}_{+}\cdot[\delta_0\delta_{+}(\bar{C}_{z}\bar{\rho}\bar{C}_{+}-\bar{\rho}\bar{C}_{+}\bar{C}_{z})-\delta_{+}\delta_{-}(\bar{C}_{+}\bar{\rho}\bar{C}_{+}-\bar{\rho}\bar{C}_{+}\bar{C}_{+})]\nonumber\\
&-&\Gamma^{\prime}_{-}\cdot[\delta_0\delta_{-}(\bar{C}_{z}\bar{\rho}\bar{C}_{-}-\bar{\rho}\bar{C}_{-}\bar{C}_{z})+\delta_{+}\delta_{-}(\bar{C}_{-}\bar{\rho}\bar{C}_{-}-\bar{\rho}\bar{C}_{-}\bar{C}_{-})]+\mathrm{h.c.}
\end{eqnarray}
\end{widetext}
The notation $\mathrm{h.c.}$ represents Hermitian conjugate. The first part in Eq. (9) is the unitary one. $\bar{H}_{L}$ is the Lamb shift Hamiltonian and describes a small shift in the energy of the eigenvectors of $\bar{H}_{eff}$. The Lamb shift Hamiltonian has no qualitative effect on the decoherence of the system and may be neglected \cite{Haikka}. Meanwhile, according to the calculation of the effective spin-electric coupling, the characteristic time for the low-energy molecular magnet of $Cu_{3}$ is about $\tau_{s}=\omega_{s}^{-1}\sim 10 \mathrm{ns}$ \cite{Islam} which is always much smaller than the correlation time of the thermal environment $\tau_{E}\sim 1 \mathrm{ms}$ \cite{Kuhr}. Under the condition of $\tau_{s}\ll\tau_{E}$, the influence of the last term in Eq. (9) on the decoherence of the chiral states is very small and usually negligible \cite{Breuer01}. However, it is also noticed that the dynamics of the non-Lindblad operator $NL[\bar{\rho}(t)]$ is indispensable for the decoherence under the circumstance of $\tau_{s}\gg \tau_{E}$. In the following, the parts of $\bar{H}_{eff}$ and Lindblad superoperator $L[\bar{\rho}(t)]$ are dominant in the second-order time-convolutionless master equation describing the dynamics of spin states in an electrically driven single molecular magnet.

\section{NON-MARKOVIAN DYNAMICS OF SPIN STATES}

An efficient way to describe the dynamics of the low-energy spin chiral states is to analyze the Bloch vector $\vec{R}(t)=(R_x,R_y,R_z)$ for the reduced density matrix $\bar{\rho}(t)$ of the single molecular magnet. The three components of the Bloch vector are defined as
\begin{equation}
\label{eq:13}
R_j=\mathrm{Tr}[\bar{\rho}(t)\bar{C}_j],(j=x,y,z).
\end{equation}
For any initial state $\vec{R}(0)=(\sin \theta \cos \phi,\sin \theta \sin \phi, \cos \theta)$, the analytical expression of the Bloch vector at time $t$ are obtained as
\begin{eqnarray}
\label{eq:14}
R_{x}(t)&=&e^{-r(t)}\cdot \cos(\omega_{s}t+\phi)\sin \theta \nonumber \\
R_{y}(t)&=&e^{-r(t)}\cdot \sin(\omega_{s}t+\phi)\sin \theta \nonumber \\
R_{z}(t)&=&e^{-p(t)}\cdot \{ \cos \theta +\int_{0}^{t}dt_{1}\cdot e^{p(t_1)}[\gamma_{+}(t_1)-\gamma_{-}(t_1)] \},
\end{eqnarray}
where $r(t)=\frac 12 \int_{0}^{t}dt_{1}[\gamma_{+}(t_1)+\gamma_{-}(t_1)+4\gamma_{z}(t_1)]$ and $p(t)=\int_{0}^{t}dt_{1}[\gamma_{+}(t_1)+\gamma_{-}(t_1)]$. The angles satisfy that $\theta\in [0,\pi]$ and $\phi\in [0,2\pi]$. The traditional Lorentzian spectral density function is used to describe the thermal bosonic environment such as the quantized electromagnetic field inside a cavity. The weak interactions between the system and thermal environment are given by the spectral density function, $\sum_{j}|g_j|^2\rightarrow \int_0^{\infty}J(\omega^{\prime})d\omega^{\prime}$ where $J(\omega^{\prime})=\frac {u^2\lambda^2}{2\pi[(\omega^{\prime}-\omega_{0})^2+\lambda^2]}$. The weak coupling constant $u^2\ll \omega_{s}$ has frequency dimension. The correlation time scale is obtained as $\tau_{E}=\lambda^{-1}$ where $\lambda$ denotes the width of the distribution quantifying leakage of photons. $\omega_{0}$ is the center frequency of the electromagnetic field in the cavity.

The time evolution of the decay rates $\gamma_{m}(t),(m=\pm,z)$ is plotted in Fig. 1 for different temperatures. The $\gamma_{m}(t),(m=\pm,z)$ is plotted in Figs. 1(a), 1(b) and 1(c) for small value of $\frac {\omega_0-\omega}{\lambda}=0.1$. The values of $\gamma_{\pm}$ always oscillate between some positive and negative values with slightly damping rate. The amplitude of $\gamma_{\pm}(t)$ for $T=0$ is much smaller than that for $T=1$. In non-Markovian quantum jumps formalism, negative values are regarded as the occurrence of reversed quantum jumps which can indicate the non-Markovian dynamics induced by the environmental memory. The memory effects describe the exchange of energy and information between the system and environment. The $\gamma_{m}(t),(m=\pm,z)$ is plotted in Figs. 1(d), 1(e) and 1(f) for large value of $\frac {\omega_0-\omega}{\lambda}=10$. The curve of $T=0$ is almost the same as that of $T=1$. That is, the influence of low temperatures on the dynamics of the decay rates is almost negligible. This is because that larger values of $\frac {\omega_0-\omega}{\lambda}$ represent smaller effective coupling between the system and environment. Meanwhile, the effects of temperatures are also very weak. However, for a small value of $\frac {\omega_0-\omega}{\lambda}$ as shown in Figs. 1(a), 1(b) and 1(c), the oscillation of the decay rates at low temperature is dramatic in comparison with that of $T=0$. It is seen that the impact of $\gamma_{z}$ on the decoherence is dominant in the the condition of $\tau_{s}\ll \tau_{E}$. For large value of $\frac {\omega_0-\omega}{\lambda}$, the non-Markovian oscillation of $\gamma_{z}$ occurs due to the negative values of $\gamma_{z}$. For long time limit, the values of the decay rate $\gamma_z$ gradually approach to some steady values close to zero. This means that the non-Markovian dynamics on the correlation time scale is reduced to the Markovian limit on the long time scale.

The $z$-component of Bloch vector $R_z$ is plotted in Fig. 2 as a function of $\lambda t$. The phenomenon of the collapses and revivals of $R_z$ indicates that the rapid exchange of energy and information happens between the system and environment. It is shown that the decoherence of spin states can be suppressed to a certain extent when the value of $\frac {\omega_0-\omega}{\lambda}$ is large. To enhance the coherence of low-energy spin states, the frequency $\omega$ of the external electric field and the width $\lambda$ of spectral density function can be decreased. Fig. 3 also demonstrates the effects of the electric field on the decoherence. For larger values of $\frac {\Delta_{so}}{\omega_{s}}$, the decoherence can also be restrained. It is found out that the coherence can be improved by the manipulation of the frequency $\omega$ of the electric field when $\omega$ is almost resonant with the transition frequency $\omega_{so}$ of the spin-orbit interaction. The increase of the strength of the electric field is also useful for the control of the decoherence.

In respect to quantum information processing, the stability of the information storage needs to be analyzed when encoding qubits in single molecular magnet are electric-controllable in the thermal environment. From the perspective of von Neumann entropy, the pointer state \cite{Paz, Khodjastech} can be defined as one initial state which becomes minimally entangled with the environment during the evolution. The study of the pointer state can help us to understand the effects of the decoherence on quantum information processing. The entropy for the reduced density matrix $\bar{\rho}(t)$ of the non-Markovian decoherence can be written as
\begin{equation}
\label{eq:15}
E(t)=-\mathrm{Tr}[\bar{\rho}(t)\ln\bar{\rho}(t)]=\sum_{i=1,2}v_{i}\ln v_{i},
\end{equation}
where the eigenvalues of $\bar{\rho}$ are given by $v_{i}=\frac {1\pm \sqrt{\sum_j |R_j|^2}}{2}$. By means of Eq. (14), the entropy $E(t)$ is easily calculated and dependent on the angle $\theta$ of the initial spin state.

For zero temperature of $T=0$, the entropy $E$ during the decoherence is plotted in Figs. 4. It is seen that the values of the entropy are always increased with the time in Fig. 4(a). The larger values of the entropy denote the more entanglement between the system and environment. Moreover, the time-dependent behavior of the entropy also shows the non-Markovian oscillation in Fig. 4(b). The revivals of the entropy are induced by the environmental memory effects. When the initial state is at $\theta=\pi$, the values of $E$ almost remain the minimal ones in the evolution. Therefore, the pointer state for the thermal environment with $T=0$ is the state of $\theta=\theta_{p}=\pi$ which is almost the ground state $|\psi_{-}\rangle$. The effects of the environment on the pointer state at $T=0$ is very weak. In fact, the selection of the pointer state is determined by the properties of the environment \cite{Paz, Khodjastech}. At low temperature of $T=1$, the dynamics of the entropy is shown in Fig. 5. From Fig. 5(a), it is clearly seen that the values of the entropy for $T=1$ are always increased more quickly than those of $T=0$. This means that the single molecular magnet is easily entangled with the environment at $T\neq 0$. It is found out that the pointer state with the initial angle $\theta_{p}$ are varied with the parameter $\frac {\omega_0-\omega}{\lambda}$. For large value of $\frac {\omega_0-\omega}{\lambda}\geq 10$, the pointer state is approximately the ground state because of the very small effective coupling between the system and the environment.

\section{DISCUSSION}

The decoherence of the low-energy spin states in an electrically driven single molecular magnet weakly coupled to a thermal environment is investigated. By means of the time-convolutionless non-Markovian master equation, the reduced density matrix for the spin states can be derived in the condition of $\tau_{s}\ll \tau_{E}$. In regard to the Lorentzian environment, the oscillations of the decay rates between positive values and negative ones appear. This phenomenon indicates the memory effects from the non-Markovian environment. The rapid non-Markovian decoherence of the Bloch vector occurs due to the quick exchange of energy and information between the system and the environment. The decoherence can be efficiently suppressed by adjusting the electric field and the parameters of the environmental spectral density function. In quantum information processing, the selection of the pointer states can be determined by the properties of the environment.

\section{ACKNOWLEDGEMENT}

This work is supported by the National Natural Science Foundation of China under Grant No. 10904104 and No. 11074184.

\newpage

{\Large \bf Figure Captions}

{\bf Figs. 1}

The time evolution of the decay rates $\gamma_{m}(t),(m=\pm,z)$ is plotted as a function of the scaled time $\lambda t$ when the parameters are $\frac {\omega_{s}}{\lambda}=100$ and $\frac {\Delta_{so}}{\omega_{s}}=0.4$. For (a), (b) and (c), the parameter is $\frac {\omega_0-\omega}{\lambda}=0.1$ while for (d), (e) and (f), the parameter is $\frac {\omega_0-\omega}{\lambda}=10$. The black solid line denotes the case of $T=1$ and red solid one represents that of $T=0$.

{\bf Fig. 2}

The dynamics of the z component of the Bloch vector $R_{z}$ is plotted as a function of the scaled time $\lambda t$ for $T=1$, $\frac {\omega_{s}}{\lambda}=100$ and $\frac {\Delta_{so}}{\omega_{s}}=0.4$. The black solid line denotes the case of $\frac {\omega_0-\omega}{\lambda}=10$ and the red one represents that of $\frac {\omega_0-\omega}{\lambda}=0.1$.

{\bf Fig. 3}

The dependence of the decoherence of the spin states on the electric field is plotted for $\frac {\omega_{s}}{\lambda}=100$, $\frac {\omega_0-\omega}{\lambda}=0.1$ and $T=1$. The parameter $\frac {\Delta_{so}}{\omega_{s}}$ can be modified with the frequency $\omega$ or the strength $\epsilon$ of the electric field.

{\bf Figs. 4}

The dynamics of the von Neumann entropy $E$ is plotted at $T=0$. (a) The initial state are changed with $\theta$ and the parameters are $\frac {\omega_{s}}{\lambda}=100$, $\frac {\omega_0-\omega}{\lambda}=0.1$ and $\frac {\Delta_{so}}{\omega_{s}}=0.9$. (b) The $E$ is plotted for $\theta=0$ (black line), $\theta=\frac {\pi}{2}$ (red line) and $\theta=\pi$ (green line). The non-Markovian dynamics of the entropy is clearly presented.

{\bf Figs. 5}

The dynamics of the von Neumann entropy at $T=1$ is plotted for $\frac {\omega_{s}}{\lambda}=100$ and $\frac {\Delta_{so}}{\omega_{s}}=0.9$. (a) The initial state are changed with $\theta$ when $\frac {\omega_0-\omega}{\lambda}=10$. (b) The pointer state represented by $\theta_{p}$ is plotted as a function of the environment parameter $\frac {\omega_0-\omega}{\lambda}$.

\newpage

\begin{figure}
\includegraphics{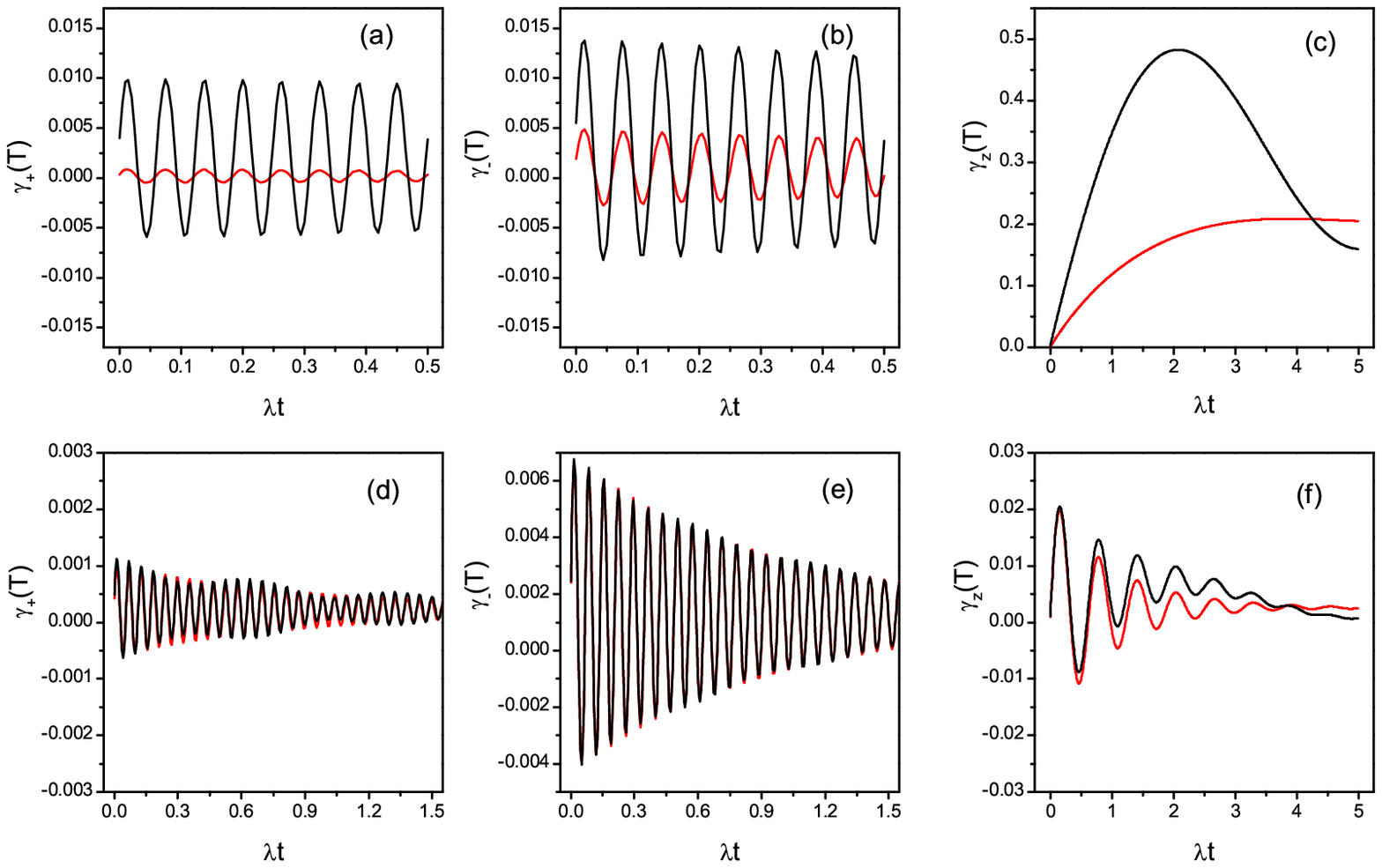}
\end{figure}

\begin{figure}
\includegraphics{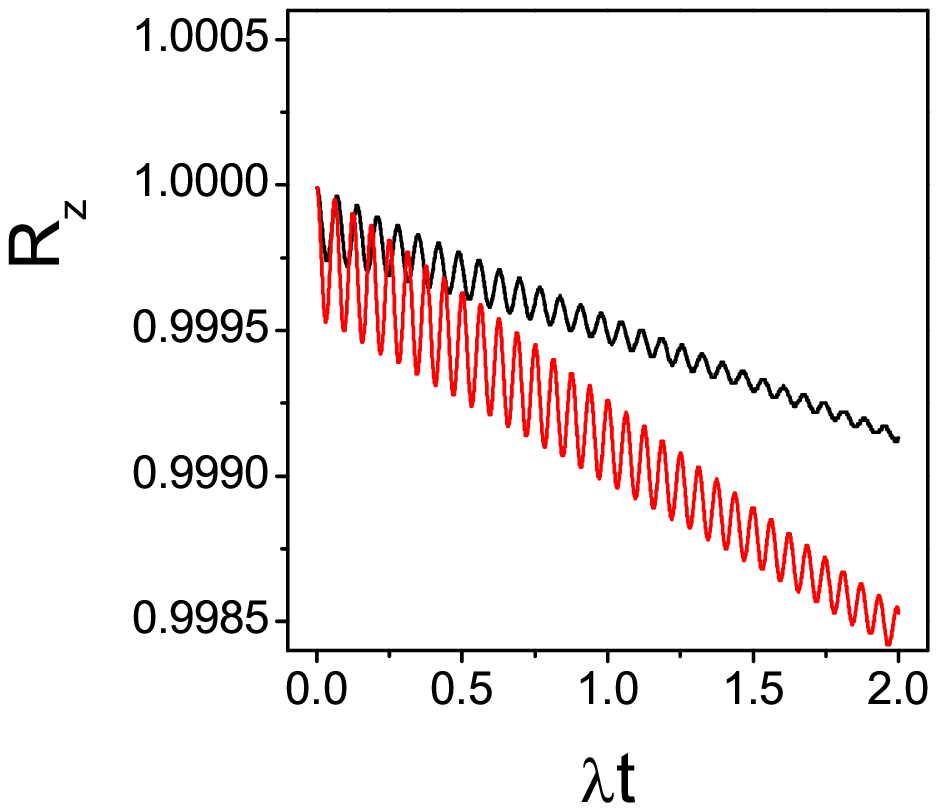}
\end{figure}

\begin{figure}
\includegraphics{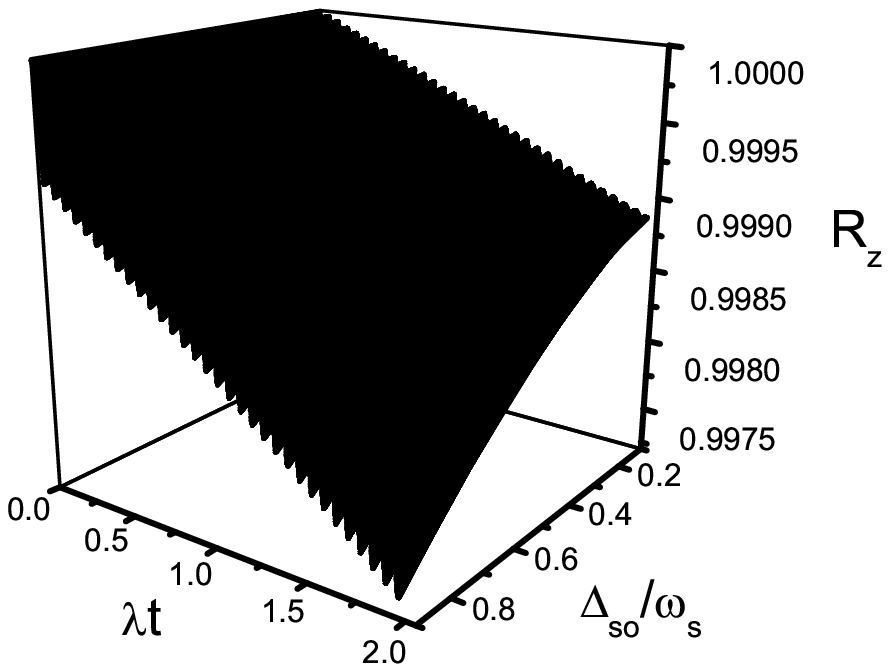}
\end{figure}

\begin{figure}
\includegraphics{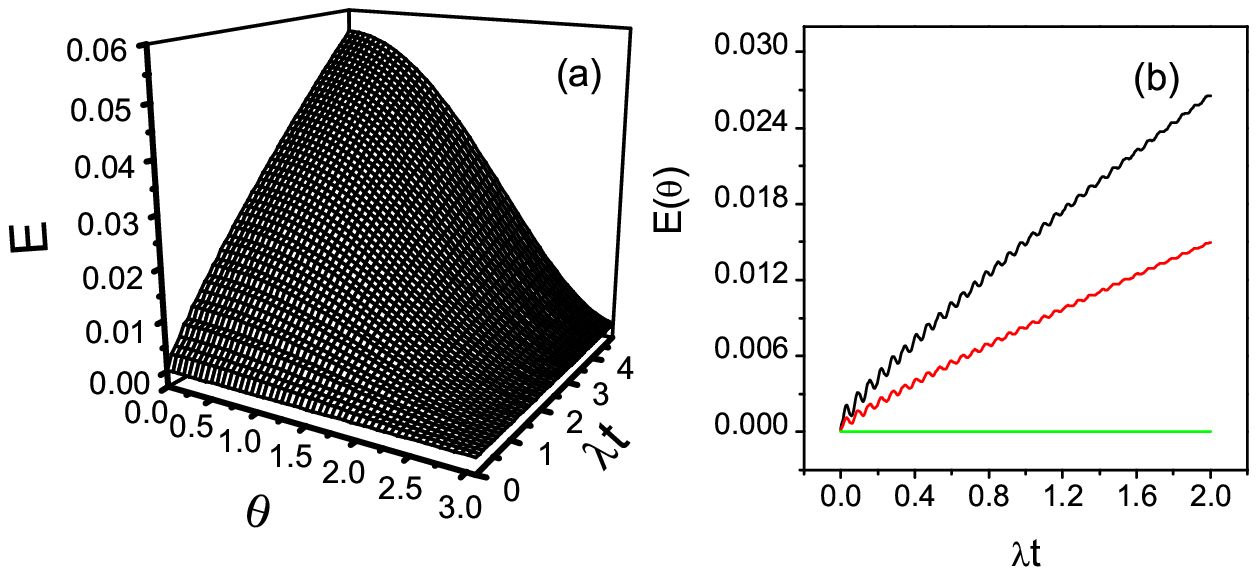}
\end{figure}

\begin{figure}
\includegraphics{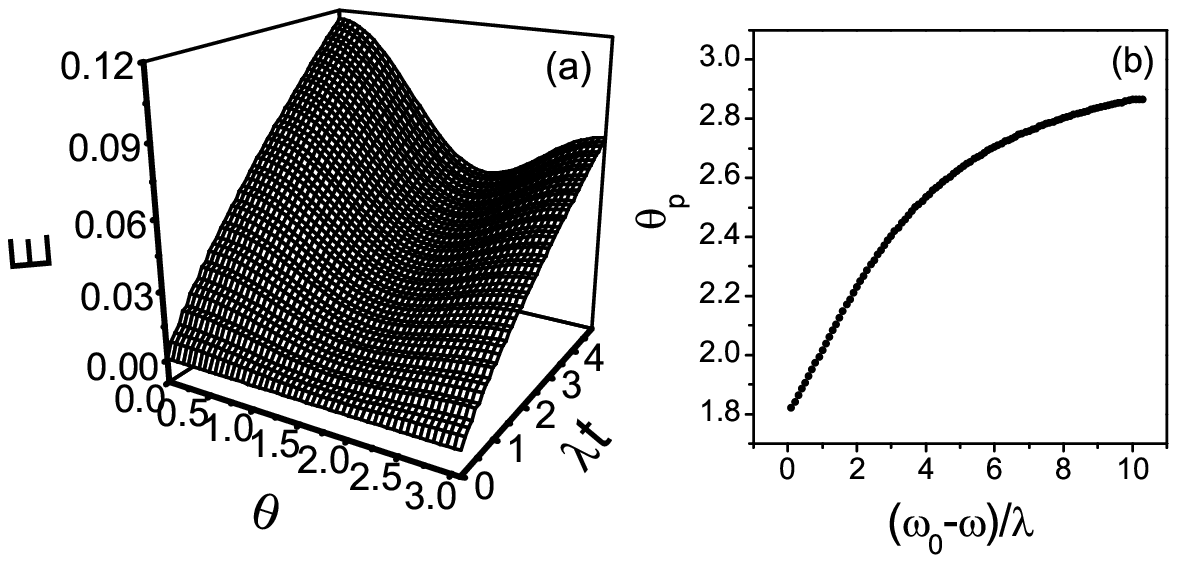}
\end{figure}


\begin{references}
\bibitem{Breuer01} H.-P. Breuer and F. Petruccione, \textit{The Theory of Open Quantum Systems}(Oxford University Press, Oxford, 2001)
\bibitem{Weiss99} U. Weiss, \textit{Quantum Dissipative Systems}(World Scientific, Singapore, 1999)
\bibitem{Hoeppe} U. Hoeppe, C. Wolff, J. K\"{u}chenmeister, \textit{et al.}, Phys. Rev. Lett. \textbf{108}, 043603(2012).
\bibitem{Guo} J.-S. Xu, C.-F. Li, \textit{et al.}, Phys. Rev. A \textbf{82}, 042328(2010).
\bibitem{Kuhr} S. Kuhr, \textit{et al.}, Appl. Phys. Lett. \textbf{90}, 164101(2007).
\bibitem{Carle} T. Carle, H.J. Briegel, and B. Kraus, Phys. Rev. A \textbf{84}, 012105(2011).
\bibitem{Ardavan} A. Ardavan, O. Rival, J.J.L. Morton, \textit{et al}., Phys. Rev. Lett. \textbf{98}, 057201(2007).
\bibitem{Gatteschi} D. Gatteschi, R. Sessoli, and J. Villain, \textit{Molecular Nanomagnets}(Oxford University Press, Oxford, 2006)
\bibitem{Friedman} J. R. Friedman, M. P. Sarachik, J. Tejada, and R. Ziolo,Phys. Rev. Lett. \textbf{76}, 3830(1996).
\bibitem{Thomas} L. Thomas, F. Lionti, R. Ballou, \textit{et al}., Nature (London) \textbf{383}, 145(1996).
\bibitem{Wernsdorfer} W. Wernsdorfer and R. Sessoli, Science \textbf{284}, 133(1999).
\bibitem{Leuenberger} M. N. Leuenberger and D. Loss, Nature(London) \textbf{410}, 789(2001).
\bibitem{Lehmann} J. Lehmann, A. Gaita-Arino, E. Coronado, and D. Loss, Nat. Nanotechnol. \textbf{2}, 312(2007).
\bibitem{Geogeot} B. Georgeot and F. Mila, Phys. Rev. Lett. \textbf{104}, 200502(2010).
\bibitem{Timco} G. A. Timco, \textit{et al}., Nat. Nanotechnol. \textbf{4}, 173(2008).
\bibitem{Candini} A. Candini, \textit{et al}., Phys. Rev. Lett. \textbf{104}, 037203(2010).
\bibitem{Szallas} A. Szallas and F. Troiani, Phys. Rev. B \textbf{82}, 224409(2010).
\bibitem{Morello} A. Morello, P.C.E. Stamp, and I. S. Tupitsyn, Phys. Rev. Lett. \textbf{97}, 207206(2006).
\bibitem{Prokofev} N. V. Proko$f^{,}$ev and P.C.E. Stamp, Rep. Prog. Phys. \textbf{66}, 669(2000).
\bibitem{Coish} W. A. Coish and J. Baugh,Phys. Status Solidi B \textbf{246}, 2203(2009).
\bibitem{Biercuk} M. J. Biercuk, H. Uys, A. P. vanDevender, \textit{et al}., Nature (London) \textbf{458}, 996(2009).
\bibitem{Maniscalco} S. Maniscalco, J. Piilo, F. Intravaia, F. Petruccione, and A. Messina Phys. Rev. A \textbf{69}, 052101(2004).
\bibitem{Haikka} P. Haikka and S. Maniscalco, Phys. Rev. A \textbf{81}, 052103(2010).
\bibitem{Goan} H.-S. Goan, P.-W. Chen, and C.-C. Jian, J. Chem. Phys. \textbf{134}, 124112(2011).
\bibitem{Troiani} F. Troiani, A. Ghirri, M. Affronte, \textit{et al}., Phys. Rev. Lett. \textbf{94}, 207208(2005).
\bibitem{Trif} M. Trif, F. Troiani, D. Stepanenko, and D. Loss, Phys. Rev. Lett. \textbf{101}, 217201(2008).
\bibitem{Hirji06} C. F. Hirjibehedin, C. P. Lutz, and A. J. Heinrich, Science \textbf{312}, 1021(2006).
\bibitem{Chiorescu} I. Chiorescu, W. Wernsdorfer, A. Muller, H. Bogge, and B. Barbara, Phys. Rev. Lett. \textbf{84}, 3454(2000).
\bibitem{Juan} M. C. Juan \textit{et al}., Inorg. Chem. \textbf{44}, 3389(2005).
\bibitem{Luzon} J. Luzon \textit{et al}., Phys. Rev. Lett. \textbf{100}, 247205(2008).
\bibitem{Trif10} M. Trif, F. Troiani, D. Stepanenko, and D. Loss, Phys. Rev. B \textbf{82}, 045429(2010).
\bibitem{Islam} M. F. Islam, J. F. Nossa, C. M. Canali, and M. Pederson, Phys. Rev. B \textbf{82}, 155446(2010).
\bibitem{Paz} J. P. Paz and W. H. Zurek, Phys. Rev. Lett. \textbf{82}, 5181(1999).
\bibitem{Khodjastech} K. Khodjasteh, V. V. Dobrovitski, and L. Viola, Phys. Rev. A \textbf{84}, 022336(2011).

\end{references}
\end{document}